\documentclass[aps,prl,twocolumn,nofootinbib,superscriptaddress]{revtex4-2}
\usepackage{graphicx}
\usepackage{amsmath}
\usepackage{epsfig}
\usepackage{amssymb}
\usepackage{xcolor}
\usepackage{cancel}
\usepackage{ulem}
\usepackage{lineno}

\begin{document}


\title{Experimental evidence of the effect of nuclear shells\\ on fission dissipation and time}

\author{D.~Ramos}
\email[]{diego.ramos@ganil.fr}
\affiliation{IGFAE - Universidade de Santiago de Compostela, E-15706 Santiago de Compostela, Spain}
\affiliation{GANIL, CNRS/IN2P3, CEA/DRF, bd Henri Becquerel, 14076 Caen, France}

\author{M.~Caama\~no}
\email[]{manuel.fresco@usc.es}
\affiliation{IGFAE - Universidade de Santiago de Compostela, E-15706 Santiago de Compostela, Spain}

\author{F.~Farget}
\affiliation{GANIL, CNRS/IN2P3, CEA/DRF, bd Henri Becquerel, 14076 Caen, France}

\author{C.~Rodr\'iguez-Tajes}
\affiliation{GANIL, CNRS/IN2P3, CEA/DRF, bd Henri Becquerel, 14076 Caen, France}

\author{A.~Lemasson}
\affiliation{GANIL, CNRS/IN2P3, CEA/DRF, bd Henri Becquerel, 14076 Caen, France}

\author{C.~Schmitt}
\altaffiliation[Present address: ]{IPHC Strasbourg, Universit\'e de Strasbourg CNRS/IN2P3, F-67037 Strasbourg Cedex 2, France}
\affiliation{GANIL, CNRS/IN2P3, CEA/DRF, bd Henri Becquerel, 14076 Caen, France}

\author{L.~Audouin}
\affiliation{IJC Lab, Universit\'e Paris-Saclay, CNRS/IN2P3, F-91405 Orsay Cedex, France}

\author{J.~Benlliure}
\affiliation{IGFAE - Universidade de Santiago de Compostela, E-15706 Santiago de Compostela, Spain}

\author{E.~Casarejos}
\altaffiliation[Present address: ]{CINTECX, Universidade de Vigo, E-36310 Vigo, Spain}
\affiliation{Universidade de Vigo, E-36310 Vigo, Spain}

\author{E.~Clement}
\affiliation{GANIL, CNRS/IN2P3, CEA/DRF, bd Henri Becquerel, 14076 Caen, France}

\author{D.~Cortina}
\affiliation{IGFAE - Universidade de Santiago de Compostela, E-15706 Santiago de Compostela, Spain}

\author{O.~Delaune}
\altaffiliation[Present address: ]{CEA/DAM/DIF, F-91297 Arpajon, France}
\affiliation{GANIL, CNRS/IN2P3, CEA/DRF, bd Henri Becquerel, 14076 Caen, France}

\author{X.~Derkx}
\affiliation{GANIL, CNRS/IN2P3, CEA/DRF, bd Henri Becquerel, 14076 Caen, France}

\author{A.~Dijon}
\affiliation{GANIL, CNRS/IN2P3, CEA/DRF, bd Henri Becquerel, 14076 Caen, France}

\author{D.~Dor\'e}
\affiliation{CEA Saclay, DMS/IRFU/SPhN, 91191 Gif-sur-Yvette Cedex, France}

\author{B.~Fern\'andez-Dom\'inguez}
\affiliation{IGFAE - Universidade de Santiago de Compostela, E-15706 Santiago de Compostela, Spain}

\author{G.~de~France}
\affiliation{GANIL, CNRS/IN2P3, CEA/DRF, bd Henri Becquerel, 14076 Caen, France}

\author{A.~Heinz}
\affiliation{Chalmers University of Technology, SE-41296 G\"oteborg, Sweden}

\author{B.~Jacquot}
\affiliation{GANIL, CNRS/IN2P3, CEA/DRF, bd Henri Becquerel, 14076 Caen, France}

\author{C.~Paradela}
\altaffiliation[Present address: ]{EC-JRC, Institute for Reference Materials and Measurements, Retieseweg 1111, B-2440 Geel, Belgium}
\affiliation{IGFAE - Universidade de Santiago de Compostela, E-15706 Santiago de Compostela, Spain}

\author{M.~Rejmund}
\affiliation{GANIL, CNRS/IN2P3, CEA/DRF, bd Henri Becquerel, 14076 Caen, France}

\author{T.~Roger}
\affiliation{GANIL, CNRS/IN2P3, CEA/DRF, bd Henri Becquerel, 14076 Caen, France}

\author{M.-D.~Salsac}
\affiliation{CEA Saclay, DMS/IRFU/SPhN, 91191 Gif-sur-Yvette Cedex, France}

\date{\today}

\begin{abstract}
Nuclear fission is still one of the most complex physical processes we can observe in nature due to the interplay of macroscopic and microscopic nuclear properties that decide the result. An example of this coupling is the presence of nuclear dissipation as an important ingredient that contributes to drive the dynamics and has a clear impact on the time of the process. However, different theoretical interpretations, and scarce experimental data make it poorly understood. In this letter, we present the first experimental determination of the dissipation energy in fission as a function of the fragment split, for three different fissioning systems. The amount of dissipation was obtained through the measurement of the relative production of fragments with even and odd atomic numbers with respect to different initial fission energies. The results reveal a clear effect of particular nuclear shells on the dissipation and fission dynamics. In addition, the relative production of fragments with even and odd atomic numbers appears as a potential contributor to the long-standing problem of the time scale in fission.
\end{abstract}

\maketitle

\textit{Introduction.}–– More than 80 years after its discovery~\cite{hahN39,meiN39}, nuclear fission is still one of the most challenging reactions we can study in the laboratory. While some of its basic properties can be derived by describing the fissioning nucleus in a macroscopic fashion~\cite{bohPR39}, it was soon realised that nuclear structure and single-particle excitations may also have an impact on the process~\cite{kraP40,hilPR53}. Since then, the interplay between collective and intrinsic degrees of freedom is yet to be fully understood and accounted for in current theoretical fission models~\cite{benJPG20}.

An expected effect of such an interplay are dissipative processes, where part of the energy stored in collective excitations is transferred to single-particle excitations. Nuclear dissipation is a pivotal ingredient that can be found in different phenomena, such as nuclear fusion and the production of superheavy elements~\cite{armRPP99,ariPRC99}, deep inelastic collisions~\cite{swiPS81,winNPA95}, or giant resonances~\cite{nixPRC80}. In all of them, nuclear dissipation contributes to define the time scale and outcome. The case of nuclear fission should not be different: as the system evolves from an equilibrated system at the barrier~\cite{wigner} to the scission point, intrinsic excitations draw energy from collective deformation, slowing down the process (see early discussions in~\cite{davPRC76,ledNPA77}). The actual role of dissipation and its link with the time scale in fission is still to be fully understood; different prescriptions can be found in fission models~\cite{schRPP18,bulPRC19,benJPG20,schPPNP22}, while interpretations from experimental data are usually model-dependent~\cite{hilAPF92,pauARNPS94}.

An additional consequence of the interplay between collective and intrinsic excitations would be a certain correlation between dissipation and nuclear structure. While the role of closed shells in the fissioning system was studied in the past~\cite{newNPA88,bacPRC99,sinPRC12}, few descriptions explore this dependence with the structure of the fragments~\cite{mirPRC14}, which was never observed in experimental data. However, indications of low dissipative fission were assigned in very asymmetric fragment splits~\cite{sidNPA89} and in the survival of nuclear clusters in cold fission~\cite{armRPP99}.

 \begin{figure*}[!t]
 \includegraphics[width=0.67\columnwidth]{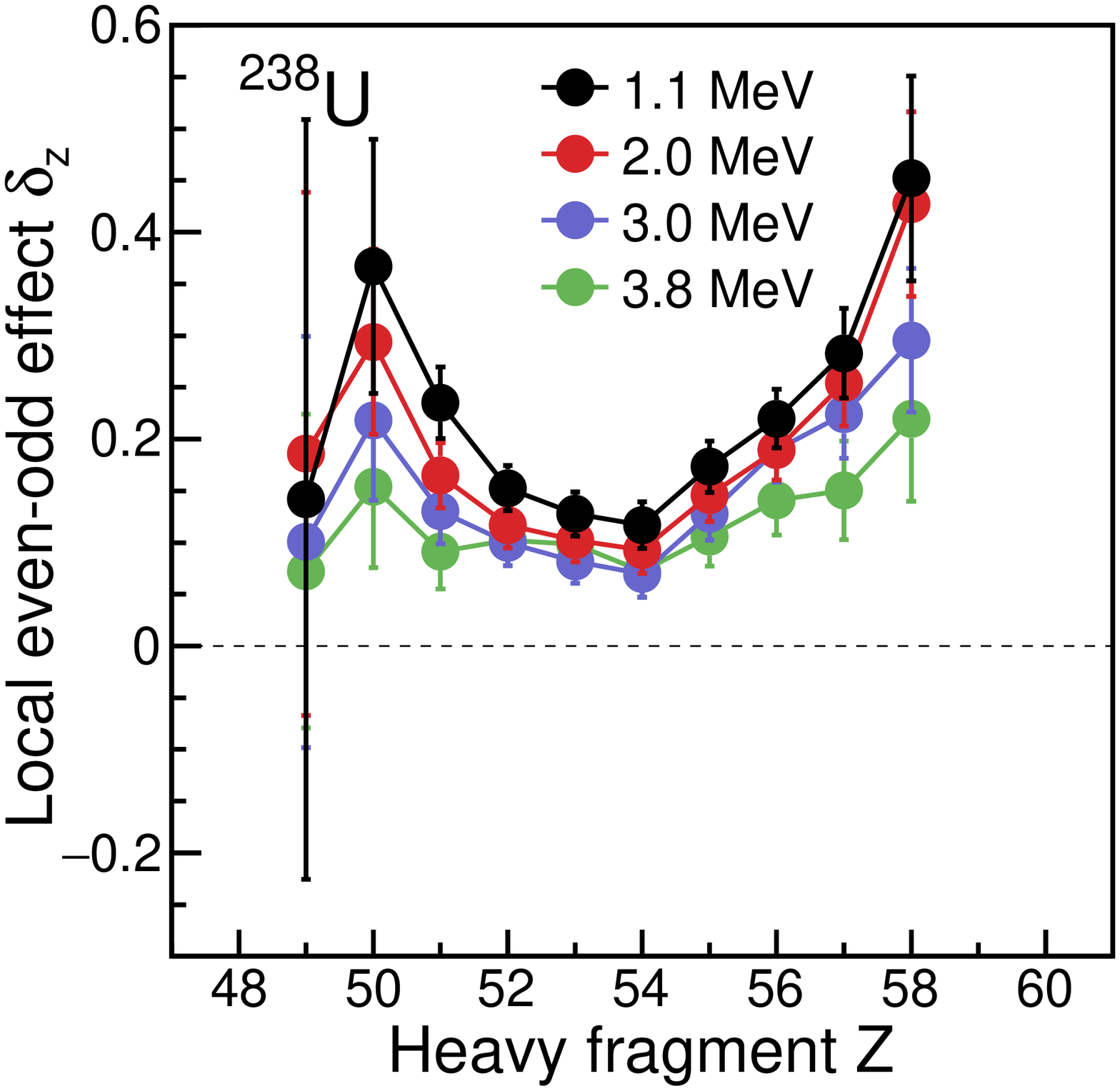}%
  \includegraphics[width=0.67\columnwidth]{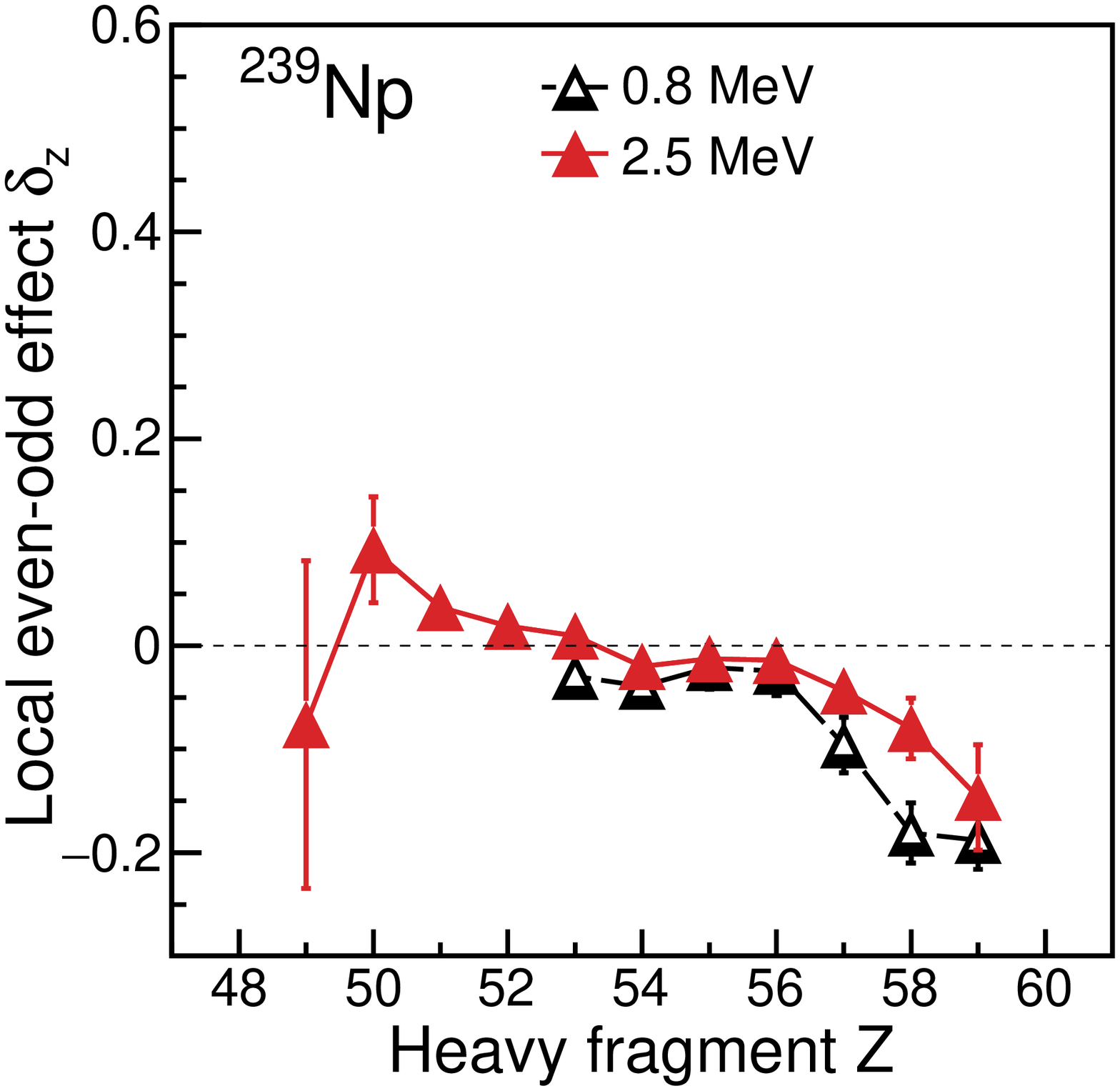}%
  \includegraphics[width=0.67\columnwidth]{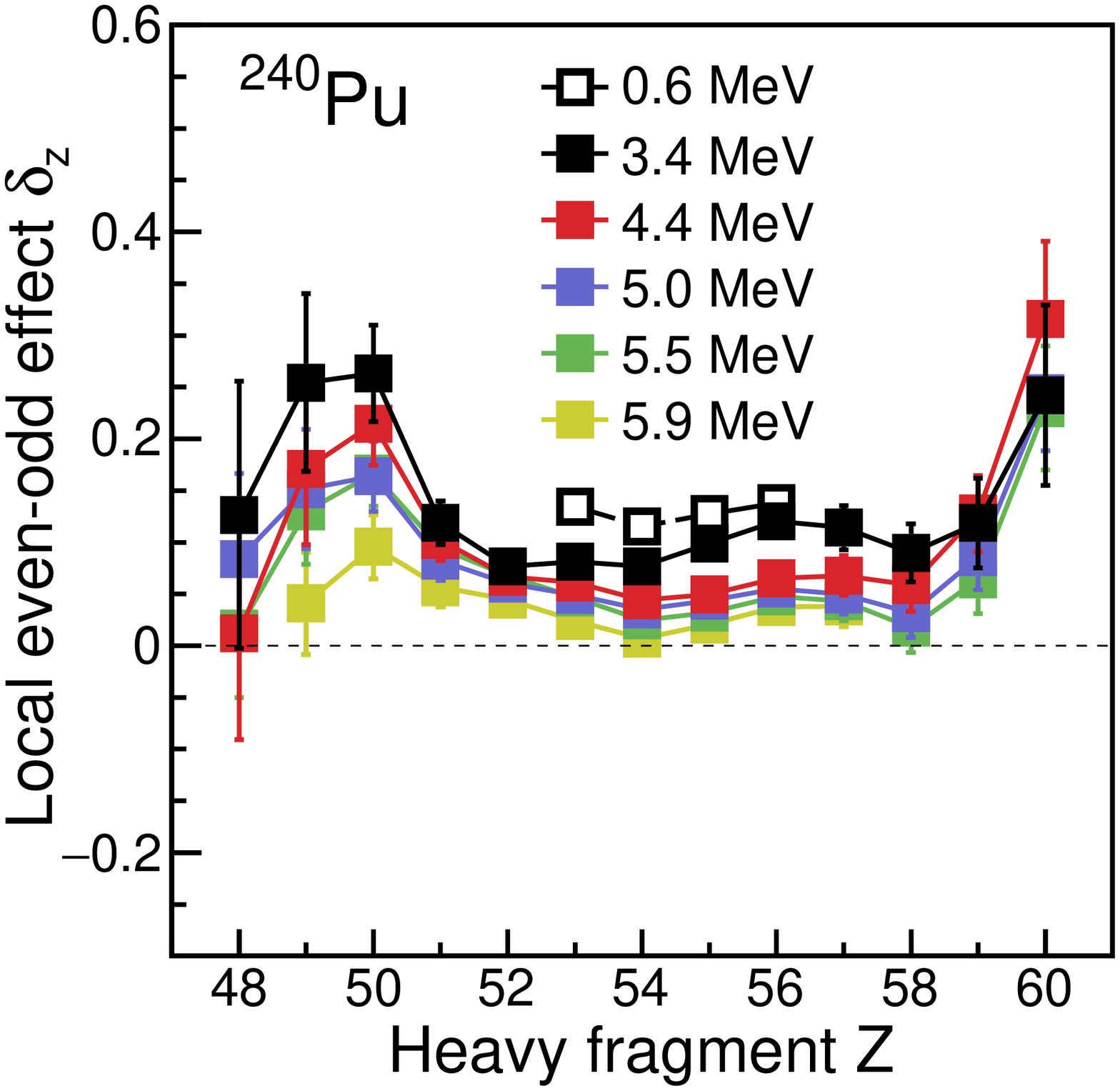}
  \caption{Local even-odd effect $\delta_{\rm Z}$ as a function of the heavy-fragment Z for $^{238}$U (left), $^{239}$Np (middle) with data from~\cite{tse} (dashed line with empty symbols), and $^{240}$Pu (right) with data from~\cite{schnpa} (dashed line with empty symbols).\footnote{We only include values of $\delta_{\rm Z}$ calculated with complete isotopic distributions~\cite{rejGANIL}.} Each line and colour correspond to the $\langle E^{\rm Bf}\rangle$ listed in the figures.}
\label{f1}
\end{figure*}

Experimentally, the amount of intrinsic energy produced in fission events can be related with the global \mbox{even-odd} effect $\delta$, which is defined as the relative difference between the total yield of even-Z fragments and that of odd-Z fragments, being Z the number of protons. An example of this relation can be found in fission reactions of even-Z systems with a low initial excitation energy $E^{*}$, of the order of the fission barrier height. On their way from the ground-state deformation to the fission barrier, the systems transform the initial energy into further deformation in order to overcome the barrier, and the single-particle excitations in the transition states at the fission barrier represent the starting conditions for the dynamical evolution towards scission. In these low-energy cases, only completely paired transition states, with no single-particle excitations, are populated~\cite{rejNPA00}. However, in such reactions, odd-Z fragments were measured~\cite{amiPRC75,amiPRC77,pomNPA,perNPA}. This means that, at some point before scission, proton pairs are broken and the single protons are distributed between the pre-fragments,\footnote{Pre-fragments are still joined through the neck but they already display basic individual properties~\cite{mosPRC71,marZP72}.} resulting in odd-Z combinations (an equivalent process happens with neutrons, although neutron evaporation from fragments prevents its direct measurement). The appearance of these single protons in even-Z systems that have started their descend from saddle to scission in completely paired configurations is an evidence that part of the potential-energy release is dissipated into intrinsic excitations~\mbox{\cite{mirPLB09,simPRC14,jurJPG15}}.

The amplitude of $\delta$ was found to decrease exponentially with $E^{\rm Bf}$, the energy above the barrier $Bf$: \mbox{$E^{\rm Bf}=E^{*}-Bf$}~\cite{pomNPA,perNPA}. Since $E^{\rm Bf}$ and the dissipated energy, $E^{\rm dis}$, are intrinsic excitations, it is straightforward to assume that $\delta$ would have the same dependence on both. A combinatorial analysis led to a direct link between $E^{\rm dis}$ and $\delta$ for low-energy fission~\cite{nifZPA,gon91}:
\begin{equation}
E^{\rm dis}+E^{\rm Bf}\approx - 4 \ln(\delta).
\label{eq_gon}
\end{equation}

Similarly, it is also possible to compute the \textit{local} even-odd effect, $\delta_{\rm Z}$, which is the magnitude of the even-odd effect as a function of the fragment Z~\cite{tracy,gonNIM,olmEPJA15}. The study of $\delta_{\rm Z}$ revealed a systematic increase of the effect with the asymmetry of the split, explained with the influence of the relative level densities of the fragments~\cite{steinNPA,caaJPG}. In a recent model~\cite{jurJPG15}, $\delta_{\rm Z}$ is obtained from the statistical breaking of pairs  and their distribution between the pre-fragments, according to the energy-sorting mechanism identified in~\cite{schPRL}. However, this model does not describe explicitly the generation of the energy dissipated, assuming it as constant fraction of the potential energy gained in the process, following the prescription of~\cite{asgJPC}.

Despite the close relation between $\delta_{\rm Z}$ and the intrinsic energy, no data set on the evolution of $\delta_{\rm Z}$ with $E^{*}$ was available, so far. We present here the first set of such measurements. From their analysis, we can estimate $E^{\rm dis}$ as a function of the fragment Z, and reveal the role of spherical and deformed shells in the generation of dissipation. These results also identify different stages of energy sharing through nucleon exchange, and hint at a potential observable to compare the fission time of fragment splits.\\
 
\textit{Local even-odd effect and fission energy.}–– We have measured the local even-odd effect $\delta_{\rm Z}$ in the fragment distributions of three fissioning systems as a function of the fission excitation energy $E^*$. The data were obtained from transfer-induced fission reactions between a $^{238}$U beam and a $^{12}$C target, measured in inverse kinematics with the large-acceptance magnetic spectrometer VAMOS++ at GANIL (France)~\cite{pulNIM,rejNIM}. The main experimental details can be found in~\cite{rodPRC14,ramPRC18,ramPRC19}. In the present work, the computation of the local even-odd effect is based on the widely-used prescription from~\cite{tracy}, in which $\delta_{\rm Z}$ corresponds to the fraction that the elemental yields of groups of four consecutive fragment Z deviate from a Gaussian behaviour (see Supplemental Material~\cite{SM} for details).

The data presented here include fission from $^{238}$U, $^{239}$Np, and $^{240}$Pu, each with its corresponding measured distribution of $E^{*}$~\cite{rodPRC14} detailed in the Suplemental Material~\cite{SM}. In addition, we have included data of thermal-neutron-induced fission of $^{240}$Pu and $^{239}$Np from~\cite{schnpa,tse}. Figure~\ref{f1} shows the evolution of $\delta_{\rm Z}$ as a function of the fragment~Z and the average $E^{\rm Bf}$.

\begin{figure*}
 \includegraphics[width=2\columnwidth]{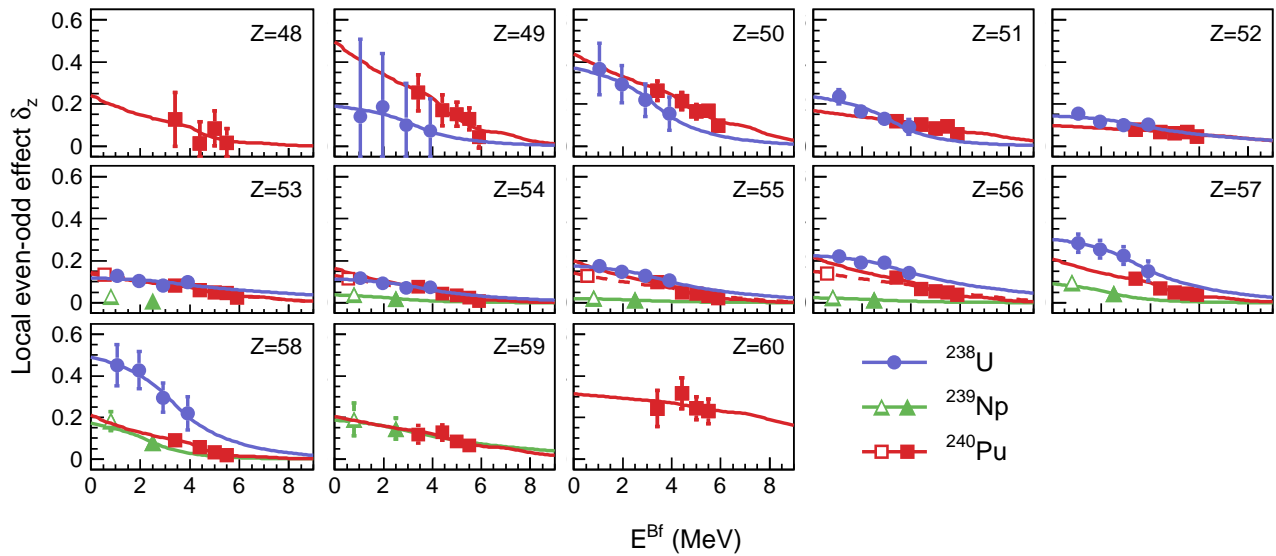}
   \caption{Evolution of $\delta_{\rm Z}$ as a function of the average $E^{\rm Bf}$ for each split in $^{238}$U (blue dots), $^{239}$Np (green triangles), and $^{240}$Pu (red squares). In the case of $^{239}$Np, $-\delta_{\rm Z}$ is plotted. Solid symbols are measured data and lines are fits to Eq.~\ref{eq_exp} folded with the measured $E^{*}$ detailed in the Suplemental Material~\cite{SM}. Empty symbols are data from~\cite{tse,schnpa}. Red dashed lines show fits when data from~\cite{schnpa} are included.}
\label{figfits}
\end{figure*}

The expected systematic increase of $\delta_{\rm Z}$ with the asymmetry of the split is observed in Fig.~\ref{f1} in $^{238}$U and $^{240}$Pu. As discussed in the previous section, unpaired protons are drawn to higher level densities, thus their probability of ending up in heavier fragments increases with the ratio of the fragment densities, closely related to the asymmetry of the split. This mechanism is clearly seen in the evolution of $\delta_{\rm Z}$ in the odd-Z system $^{239}$Np: the heavy fragment is more likely to receive the unpaired proton, resulting in negative $\delta_{\rm Z}$.

The measured $\delta_{\rm Z}$ also shows a clear maximum around the spherical closed shell Z=50 in the three systems. However, the fact that the maximum in $^{239}$Np corresponds to positive $\delta_{\rm Z}$, instead of negative values, reveals a mechanism other than the effect of level density. While it is tempting to explain it with an inversion of the energy and proton flux due to the reduced level density at Z=50, it is not supported by experimental data, which show that the energy flux towards heavy fragments is maintained along the fragment distribution~\cite{naq}. The presence of this maximum suggests that heavy pre-fragments with atomic number Z=50 are formed preferably without unpaired protons. The data from $^{239}$Np shows then two ways in which intrinsic energy contributes to $\delta_{\rm Z}$ in two different time frames: with the distribution of unpaired protons at the very formation of pre-fragments close to the barrier and with the transfer of unpaired protons between pre-fragments right until scission.

Figure~\ref{f1} also hints at the relation between the intrinsic energy and the magnitude of $\delta_{\rm Z}$: as the initial $E^{\rm Bf}$ increases, $|\delta_{\rm Z}|$ decreases. The total intrinsic energy available is the sum of the initial energy above the barrier and the dissipation generated in the process, $E^{\rm int}=E^{\rm Bf}+E^{\rm dis}$. In order to describe the dependence of $\delta_{\rm Z}$ with $E^{\rm int}$, we use a phenomenological generalisation of the prescription discussed in Eq.~\ref{eq_gon}:

\begin{align}
E^{\rm dis}(Z)&+[E^{\rm Bf}-\Delta] = G(Z)\ln{(|\delta_{\rm Z}|)} & E^{\rm Bf}>\Delta, \label{eq_exp}\\
E^{\rm dis}(Z)&= G(Z)\ln{(|\delta_{\rm Z}|)} & E^{\rm Bf}\leq\Delta. \nonumber
\label{eq_exp}
\end{align}

Here, the minimum energy needed to break proton pairs\footnote{Similar to the BCS formalism, $\Delta$ is twice the energy to produce a quasi-particle, i.e. a free proton (see Suplemental Material~\cite{SM} for details).}, $\Delta$, is subtracted from $E^{\rm Bf}$ (see Suplemental Material~\cite{SM} for details). Below this value, pairs are only broken through dissipation. This formula has two free parameters that depend on the fragment split: $E^{\rm dis}(Z)$ and $G(Z)$. The former corresponds to the intrinsic energy for initial $E^{\rm Bf}$ below $\Delta$. Such intrinsic energy is necessarily released during the process and thus, identified as dissipation energy. The parameter $G$ corresponds to the slope of the correlation between $E^{\rm Bf}$ and $\ln{(|\delta_{\rm Z}|)}$. In other words, $G$ can be interpreted as the sensitivity of the $\delta_{\rm Z}$ to changes in excitation energy. As a reference, $G$=$-$4~MeV in Eq.~\ref{eq_gon} for the case of the global even-odd effect $\delta$~\cite{gon91}.

\begin{figure*}[t]
 \includegraphics[width=\columnwidth]{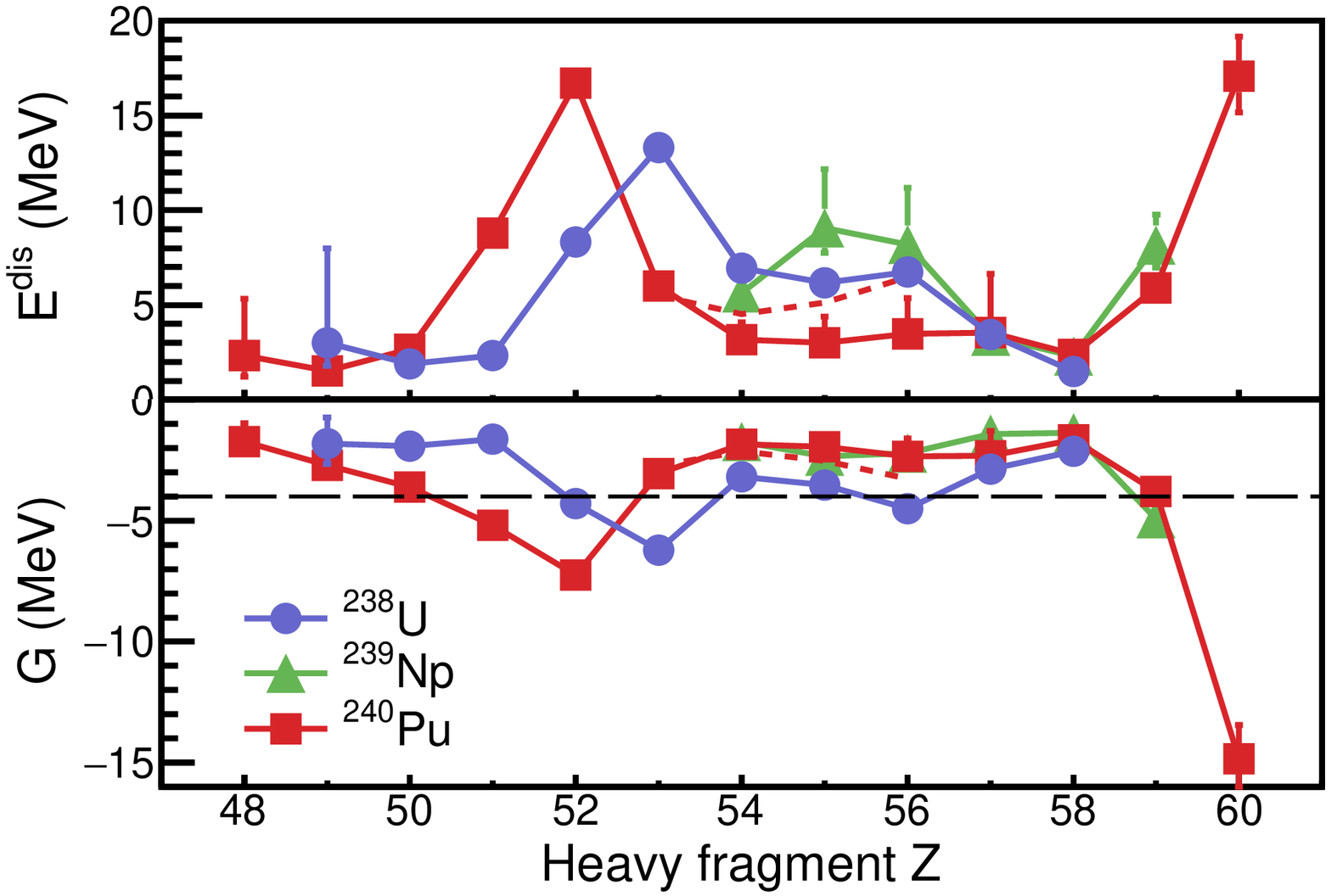}
 \includegraphics[width=\columnwidth]{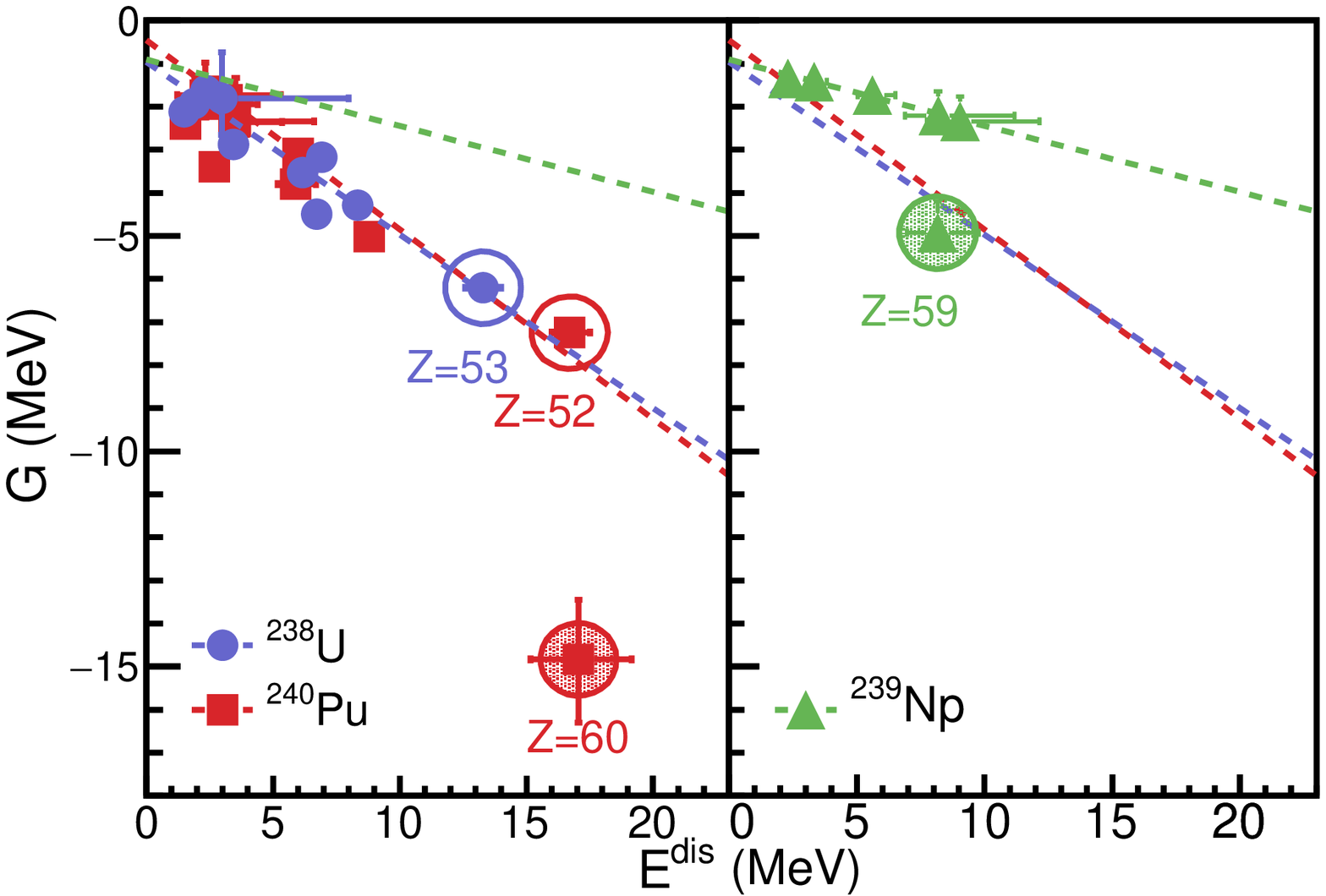}
  \caption{Left figure: Deduced $E^{\rm dis}$ (top) and $G$ (bottom) as a function of the heavy-fragment Z. The red dashed lines show the results of $^{240}$Pu when including data from~\cite{schnpa}. In the bottom panel, the horizontal dashed line shows $G$=$-4$ MeV (see text). Right figure: Correlation between $G$ and $E^{\rm dis}$ for $^{238}$U and $^{240}$Pu (left panel), and $^{239}$Np (right panel). The lines are fits to guide the eye. Circled points correspond to splits where the influence of the Z=52 shell was identified (empty enclosing), and to threshold splits (shaded enclosing). In all panels, blue dots correspond to $^{238}$U, green triangles to $^{239}$Np, and red squares to $^{240}$Pu.}
\label{par_slope}
\end{figure*}

In Eq.~\ref{eq_exp}, as in previous works, it is assumed that $E^{\rm dis}$ has a negligible dependance on $E^{\rm Bf}$, that $\Delta$ is  constant\footnote{Following the prescription of~\cite{rejNPA00}, the change in $\Delta$ along the range of energy in this work is less than 10\%, and the effect on the fitted parameters falls to $\approx$1\%.} with respect to $E^{\rm int}$, and $Bf$ is the same for all fragment splits. In order to compare it directly with experimental data, Eq.~\ref{eq_exp} is folded with the measured probability of $E^{\rm *}$ (see Suplemental Material~\cite{SM} for details). The parameters $E^{\rm dis}$ and $G$ are obtained by fitting the resulting functions to the measured data. Figure~\ref{figfits} shows the evolution of $|\delta_{\rm Z}|$ as a function of $E^{\rm Bf}$ and the corresponding fit for each heavy-fragment~Z. In the case of $^{239}$Np and $^{240}$Pu, the data from thermal-neutron-induced fission shown in Fig.~\ref{f1}~\cite{tse,schnpa} are included. Left panel of Fig.~\ref{par_slope} displays $E^{\rm dis}(Z)$ and $G(Z)$ for the three systems explored.\\

\textit{Dissipation energy and nuclear shells.}–– Octupole-deformed shells Z=52 and 56 were recently proposed as responsible for asymmetric fission in actinides~\cite{scaN,ramPRC20}, thus it is not completely unexpected that dynamic features and dissipation may have a dependence on these shells. Indeed, the left top panel of Fig.~\ref{par_slope} shows that the deduced $E^{\rm dis}$ follows a similar pattern in $^{238}$U and $^{240}$Pu that seems to react to the shell structure of the fragments. The most prominent feature is a peak that reaches beyond 10~MeV close to the deformed shell Z=52. This maximum suggests that, in the path to form this shell and the corresponding light fragment, the system goes through a relatively large number of level-crossings where fission is slowed down~\cite{bulPRL,bulPRC19}, increasing the amount of one-body dissipation and breaking additional pairs through the Landau-Zener effect~\cite{hilPR53,mirPLB09}. Around the octupole-deformed shell Z=56, $E^{\rm dis}$ falls to $\approx$5~MeV in the three systems; an indication that they seem to endure less one-body dissipation and/or level-crossing than in the case of Z$\approx$52.

Around the spherical closed shell Z=50, $E^{\rm dis}$ drops to a minimum of $\approx$2~MeV, suggesting that most of the excitation energy gained in the process is used in deforming the light fragment, while the Z=50 heavy fragment is kept mostly spherical and with low intrinsic energy. This is consistent with the maximum of $\delta_{\rm Z}$ seen in Fig.~\ref{f1} and with the low neutron evaporation measured around Z$\approx$50 (see~\cite{capNDS16} for a review). The low $E^{\rm dis}$ also suggests that the system splits in a shorter time, which is consistent with the relation between the production of Z=50 fragments and the development of a short neck in the scission configuration of $^{240}$Pu and $^{239}$U reported in~\cite{caaPLB,ramPRC20}.

For more asymmetric splits, $^{240}$Pu and $^{239}$Np display a sudden increase of $E^{\rm dis}$, while $^{238}$U data seem to stop short of a similar increase. This behaviour is addressed later.

As it was mentioned, $E^{\rm dis}$ was estimated in previous works as a constant fraction of the potential energy gained up to scission~\cite{rejNPA00,asgJPC,jurJPG15,schGEF}, which shows a smooth evolution with the fragment split, with no dependence on specific shells~\cite{pomNPA94,caaPRC15,ramPRC20}. With a typical fraction of 0.35, this approximation gives an average $E^{\rm dis}$ around 7~MeV for $^{238}$U and 9~MeV for $^{240}$Pu, slightly larger than our results, which average to 5.4 and 6.4~MeV respectively. Another recent calculation, based on strongly damped shape evolution on potential-energy surfaces, also gives a larger value of 11.3~MeV for $^{240}$Pu~\cite{molPRC}.

The bottom left panel of Fig.~\ref{par_slope} also shows the evolution of $G$ as a function of the heavy-fragment Z. The average values are $G$$\approx$$-$3 and $-$3.4~MeV for $^{238}$U and $^{240}$Pu, respectively; not dissimilar to the $G$=$-$4~MeV estimated in~\cite{gon91}. Around the octupole-deformed Z=52, we observe a clear deviation, as in the case of $E^{\rm dis}$, that makes $\delta_{\rm Z}$ less sensitive to changes in $E^{\rm Bf}$. For more asymmetric splits, beyond Z=58 in $^{240}$Pu, the magnitude of $G$ increases nearly a factor~10, rendering $\delta_{\rm Z}$ almost constant and thus insensitive to changes in $E^{\rm Bf}$. The data on $^{238}$U and $^{239}$Np stop right before, but they follow a similar trend.\\

\textit{Threshold asymmetry in the local even-odd effect.}––As it was discussed, the energy dissipated when breaking proton pairs is reflected in the measured $\delta_{\rm Z}$ in two ways: with the redistribution of these protons during the formation of the pre-fragments, and with their transfer between pre-fragments along the process. While the former is mainly ruled by stochastic breaking and the structure of the pre-fragments, the latter is dominated by the level densities and the temperature difference between pre-fragments, which also depend on the asymmetry of the split~\cite{jurJPG15, schPP}. In addition, the transfer of protons needs a certain time, $t_{\rm tr}$, to be completed. In average, this time increases with the amount of intrinsic energy stored in the fragment and decreases with the mass asymmetry~\cite{schPP}.

The exchange of protons through the neck is possible until the Coulomb repulsion between the two nascent fragments approaching scission hinders it. If, for a sufficiently large asymmetry, the time until this point, $t_{\rm sc}$, is much longer than $t_{\rm tr}$, having more initial $E^{\rm int}$ would not affect $\delta_{\rm Z}$: the few unpaired protons that remain in the light fragment at the final stage of the energy sorting would have time enough to be all transferred~\cite{jurJPG15, schPP}. The result is that, for fragment asymmetries beyond a threshold point, the measured $\delta_{\rm Z}$ loses its sensitivity on $E^{\rm int}$, that is, the amplitude of $G$ increases very fast. The rapid evolution of $G$ for large asymmetry in $^{240}$Pu, shown in the left bottom panel of Fig.~\ref{par_slope}, suggests that the threshold point for this system is around Z=60. Assuming the same $t_{\rm sc}$~\cite{schPP} would result in similar threshold asymmetries for $^{238}$U and $^{239}$Np, which is consistent with the parallel evolution of $G$ in the three systems for very asymmetric splits. Since $t_{\rm tr}$$\approx$$t_{\rm sc}$ at the threshold asymmetry, the estimation of $t_{\rm tr}$ from microscopic models would inform directly on the magnitude of $t_{\rm sc}$ , adding another piece to the long-standing problem on the  time scale in fission.

Since there is a certain correlation between dissipation and the time to scission, for high $E^{\rm dis}$, it is more probable to transfer all protons from light to heavy pre-fragments, which, in turn, decreases the sensitivity of $\delta_{\rm Z}$ or, equivalently, increases the amplitude of $G$. This must result in a correlation between $E^{\rm dis}$ and $G$. Right panels of Fig.~\ref{par_slope} show that such a correlation appears for the three systems. $^{238}$U and $^{240}$Pu seem to have the same trend while $^{239}$Np, probably due to its odd-Z nature, follows a different but also well-defined correlation with a slope that is approximately half of the one corresponding to $^{238}$U and $^{240}$Pu. The measured data show two clear deviations from these correlations in $^{239}$Np and $^{240}$Pu at the splits corresponding to Z=59 and Z=60, respectively, already discussed as threshold asymmetries. These deviations support the interpretation that, beyond these asymmetries, the few unpaired protons stored in the light fragment have time enough to be transferred to the heavy one, degrading the sensitivity of $\delta_{\rm Z}$ to $E^{\rm int}$ and losing the underlying dependence on time. For splits with a heavy-fragment Z beyond these threshold points, the dissipation deduced with Eq.~\ref{eq_exp} is probably no longer reliable. The observation of these threshold points also indicates that the evolution of $\delta_{\rm Z}$ as a function of $E^{\rm int}$ in highly asymmetric splits is a potential observable to compare the fission time of different systems.\\

In summary, high-precision data on fission fragment yields from different fissioning systems at different excitation energies have been used to extract the evolution of the dissipation energy as a function of the fragment split. The analysis of the relative production of even- and odd-Z fragments shows a clear influence of nuclear structure in the dissipation process and fission dynamics: fission is faster when producing fragments with spherical shell Z=50, but slower and more dissipative when fragments are shaped around octupole-deformed shell Z=52.\\

\begin{acknowledgments}
This work was partially supported by the Spanish Ministry of Research and Innovation under the budget items FPA2010- 22174-C02-01 and RYC-2012-11585. The excellent support from the GANIL staff during the experiment is acknowledged.
\end{acknowledgments}

\pagebreak

\section*{\textbf{SUPPLEMENTAL MATERIAL}}

\section{Phenomenological relation between intrinsic energy and local even-odd staggering\label{equation}}

\subsection{Pairing gap\label{sec_pairing}}
The equation that relates the local even-odd staggering $\delta_{\rm Z}$ and the intrinsic energy presented in the main text characterises the experimental data with two parameters: $G$, which is the rate between the variation of the initial excitation energy above the barrier $E^{\rm Bf}$ and $\ln{(\delta_{\rm Z})}$; and the dissipation energy $E^{\rm dis}$, which corresponds to the intrinsic energy when $E^{\rm Bf}$=0. In addition, as in the BCS formalism, a pairing-gap energy $\Delta$ is included: below this value, no proton pairs are broken. It is important to note that, since this equation describes the evolution of $\delta_{\rm Z}$ with $E^{\rm Bf}$, the deduced $E^{\rm dis}$ corresponds to the total energy dissipated in intrinsic excitations of both neutrons and protons.

\begin{figure}
 \includegraphics[width=\columnwidth]{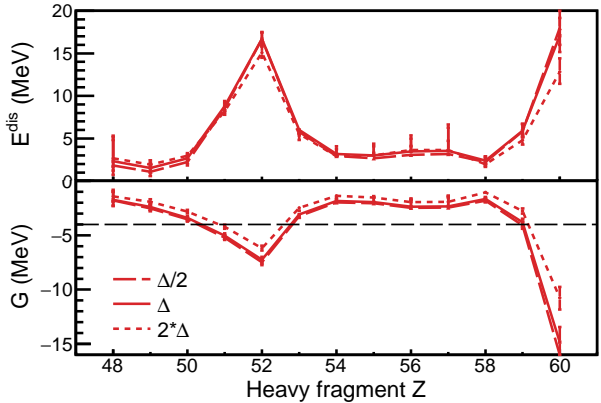}
  \caption{Resulting $E^{\rm dis}$ (top) and $G$ (bottom) for $^{240}$Pu data when fitted with a pairing gap of $\Delta/2$ (long-dashed line), $\Delta$ (solid line), and $2\Delta$ (short-dashed line).} 
\label{figD}
\end{figure}

Following the BCS formalism, the pairing-gap energy needed to break a proton pair is twice the energy needed to produce a quasi-particle. In this work, this is calculated following ref.~\cite{molADND}: $\Delta=2~(3.2\cdot {\rm Z}^{-1/3}$) MeV. While the estimation of the actual pairing-gap energy is an open question (see, for instance\cite{pomNPA,perNPA,rejNPA00,jurJPG,ivaPRC18}), our results are robust to changes in $\Delta$. Figure~\ref{figD} shows the resulting $E^{\rm dis}$ and $G$ for values of $\Delta$ between twice and half of the value used in our work.

\subsection{Excitation energy distribution\label{sec_dist_exc}}

\begin{table} [!t]
\begin{center}
  \caption{\label{table1} Fissioning systems studied in this work. The average fission barrier $Bf$ takes into account pre-fission neutron evaporation. The $E^*$ limits mark each bin in excitation energy. $E^*_c$ is the average excitation energy $E^*$ corrected for pre-fission evaporation with the \texttt{GEF} code~\cite{schGEF}. $E ^{\rm Bf}$ is the average energy above barrier. All quantities are in MeV. Data from thermal-neutron-induced fission of $^{239}$Np~\cite{tse} and $^{240}$Pu~\cite{schnpa} are also included.}
  \begin{tabular}{|c|c|c|c|c|c|}
    \hline\hline
    Fissioning~~&~Average~& $E^*$  &~Average~& $E^*_c$ &~Average\\
    system & $Bf$ & limits & $E^*$ & & $E^{\rm Bf}$\\\hline
        &&&&&\\
$^{238}$U & 5.7 & 5.0 - 8.6 & 6.8 & 6.8 & 1.1\\
~&& 6.0 - 9.6 & 7.7 & 7.7 & 2.0\\
~&& 7.0 - 10.6 & 8.7 & 8.7 & 3.0\\
~&& 8.0 - 11.6 & 9.6 & 9.5 & 3.8\\\hline
 &&&&&\\
 $^{239}$Np & 5.4 & (n$_{\rm th}$,f)~\cite{tse} & 6.2 & 6.2 & 0.8\\
~ & ~ & 6.5 - 9.5 & 7.9 & 7.9 & 2.5\\\hline
&&&&&\\
$^{240}$Pu& 5.1 & (n$_{\rm th}$,f)~\cite{schnpa} & 5.6 & 5.6 & 0.6\\
~& ~ &4.0 - 10.7 & 8.5 & 8.5 & 3.4\\
~& &7.0 - 11.8 & 9.5 & 9.5 & 4.4\\
~&&8.0 - 13.3 & 10.5 & 10.1 & 5.0\\
$^{239.9}$Pu&5.1&9.0 - 15.1& 11.5 & 10.6 & 5.5\\
 $^{239.8}$Pu&5.2&10.0 - 17.3 & 12.5 & 11.1 & 5.9 \\\hline
  \end{tabular}
\end{center}
\end{table}

The present fission data of each fissioning system were measured with certain distributions of initial excitation energy $E^*$ above their ground state~\cite{rodPRC14}. In order to study the evolution of $\delta_{\rm Z}$ with $E^*$, the experimental observables were calculated for regions of $E^*$ between the limits listed in Table~\ref{table1}. The comparison and fitting of the measurements to obtain the parameters $E^{\rm dis}$ and $G$ is done with Eq. 2 folded with the measured probability of $E^{*}$, $P(E^*)$, and the limits of each energy region:
\begin{equation}
\delta_{\rm Z}\otimes P(E^*)=\frac{\int\limits_{E^{\rm *} {\rm limits}} e^{\left[{\frac{1}{G}([E^{\rm *}-Bf - \Delta]+E^{\rm dis})}\right]}P(E^{\rm *}) {\rm d}E^{\rm *}}{\int\limits_{E^{\rm *} {\rm limits}} P(E^{\rm *}) {\rm d}E^{\rm *}}.
\label{eq_exp_2}
\end{equation}

\begin{figure}
 \includegraphics[width=\columnwidth]{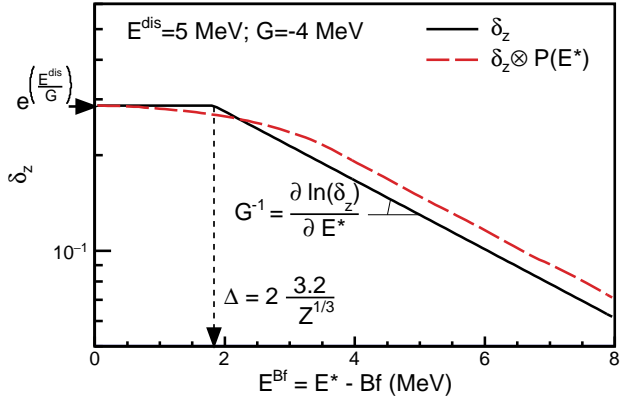}
  \caption{The black line shows the phenomenological relation between $\delta_{\rm Z}$ and $E^{\rm Bf}$, described in Eq. 2 of the main text, for the case of $^{238}$U and  typical values of $E^{\rm dis}$=5~MeV and \mbox{$G$=$-$4~MeV}. The dashed red line is the result of the folding with the experimental distribution of excitation energy described in Eq.~\ref{eq_exp_2}. Features related with the parameters $E^{\rm dis}$, $G$, and $\Delta$ are also shown.} 
\label{conv}
\end{figure}

Figure~\ref{conv} shows the formula of Eq. 2 in the main text together with the convolution of Eq.~\ref{eq_exp_2} in this document. The figure also displays some of the main features and their relation with the parameters $E^{\rm dis}$, $G$, and $\Delta$.

\section{Local even-odd staggering and Modified Tracy formula \label{sec_tracy}}
While a number of methods to calculate global and local even-odd effects can be found in the literature~\cite{gonNIM,olmEPJA15}, the formula by Tracy {\it et al.}~\cite{tracy} is probably the most commonly used for computing the local even-odd effect in nuclear fission. This method assumes that fragment yields have a constant deviation $\delta_{\rm G}$ from an underlying local Gaussian behaviour:
\begin{equation}
Y({\rm Z})=Y_{\rm G}({\rm Z})(1\pm\delta_{\rm{G}}).
\end{equation}

 By using the yields of four consecutive fragment Z, the Tracy formula calculates a unique Gaussian function and a constant deviation $\delta_{\rm Tracy}$: 
 \begin{equation}
\begin{aligned}
\delta_{\rm Tracy}({\rm Z}+1.5)&= \frac{(-1)^{{\rm Z}}}{8}\Big( \ln{Y({\rm Z})}-\ln{Y({\rm Z}+3)} \\
& + 3\big[ \ln{Y({\rm Z}+2)}-\ln{Y({\rm Z}+1)} \big]\Big).
\end{aligned}
\label{eq_tracy}
\end{equation}

The resulting $\delta_{\rm Tracy}$ is associated to the center of gravity in Z, which is 1.5 units away from the first of the series. For example, by using the yields of Z=50, 51, 52, and 53, we would obtain $\delta_{\rm Tracy}$(51.5).

\begin{figure}
 \includegraphics[width=\columnwidth]{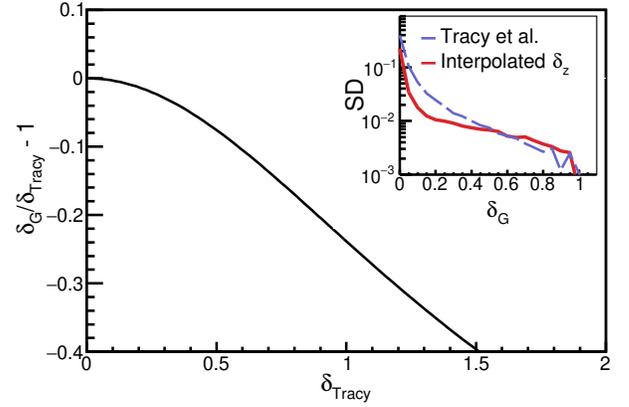}
  \caption{Main picture: Accuracy of the Tracy formula as a function of $\delta_{\rm Tracy}$. Inset: Normalised standard deviation of $\delta^{\rm corr}_{\rm Tracy}$ (dashed blue line) and $\delta_{\rm Z}$ (red solid line), as a function of the actual $\delta_{\rm{G}}$.}
\label{fig_tracy}
\end{figure}

The value of $\delta_{\rm Tracy}$ is known to deviate from $\delta_{\rm G}$ for increasing even-odd staggering~\cite{gonNIM,steinNPA}, as we can see in Fig.~\ref{fig_tracy}, where the relative difference between $\delta_{\rm G}$ and $\delta_{\rm Tracy}$ is shown. In order to correct this deviation, we calculate a corrected $\delta^{\rm corr}_{\rm Tracy}$ with a relation derived from \cite{gonNIM,steinNPA}:
\begin{equation}
\delta^{\rm corr}_{\rm Tracy}=\frac{e^{(2\delta_{\rm Tracy})}-1}{e^{(2\delta_{\rm Tracy})}+1}.
\label{eq_tracy_corr}
\end{equation}

Finally, in order to avoid the reference to semi-integer atomic numbers of the arguments in Eqs.~\ref{eq_tracy} and~\ref{eq_tracy_corr}, the even-odd effect $\delta_{\rm Z}$ used in this work is the mean value of \mbox{$\delta^{\rm corr}_{\rm Tracy}({\rm Z}-0.5)$} and \mbox{$\delta^{\rm corr}_{\rm Tracy}({\rm Z}+0.5)$}:
\begin{equation}
\delta_{\rm Z}(\rm Z)=\frac{1}{2}\left[\delta^{\rm corr}_{\rm Tracy}({\rm Z}-0.5) + \delta^{\rm corr}_{\rm Tracy}({\rm Z}+0.5)\right].
\end{equation}

This interpolation also reduces the intrinsic dispersion of the formula. We can define this dispersion, $SD$, as the normalised standard deviation of the difference between a calculated $\delta_{\rm calc}$ and the actual $\delta_{\rm{G}}$: 
\begin{equation}
SD(\delta_{\rm{G}})=\left(\sum_{i} \left( \frac{\langle \delta_{{\rm calc},i} \rangle - \delta_{{\rm calc},i} }{\delta_{\rm{G}}}\right)^2 \right)^{\frac{1}{2}}.
\end{equation}

The inset panel of Fig.~\ref{fig_tracy} shows $SD$ as a function of $\delta_{\rm{G}}$, for $\delta_{\rm calc}=\delta^{\rm corr}_{\rm Tracy}$ (dashed blue line) and for $\delta_{\rm calc}=\delta_{\rm Z}$ (red line) for a large set of different configurations of simulated fragment yields and values of even-odd staggering. Between $\delta_{\rm{G}}$=0 and 0.6, the dispersion of the interpolated value $\delta_{\rm Z}$ is lower than that of $\delta^{\rm corr}_{\rm Tracy}$.

\begin{figure*}[!t]
 \includegraphics[width=0.66\columnwidth]{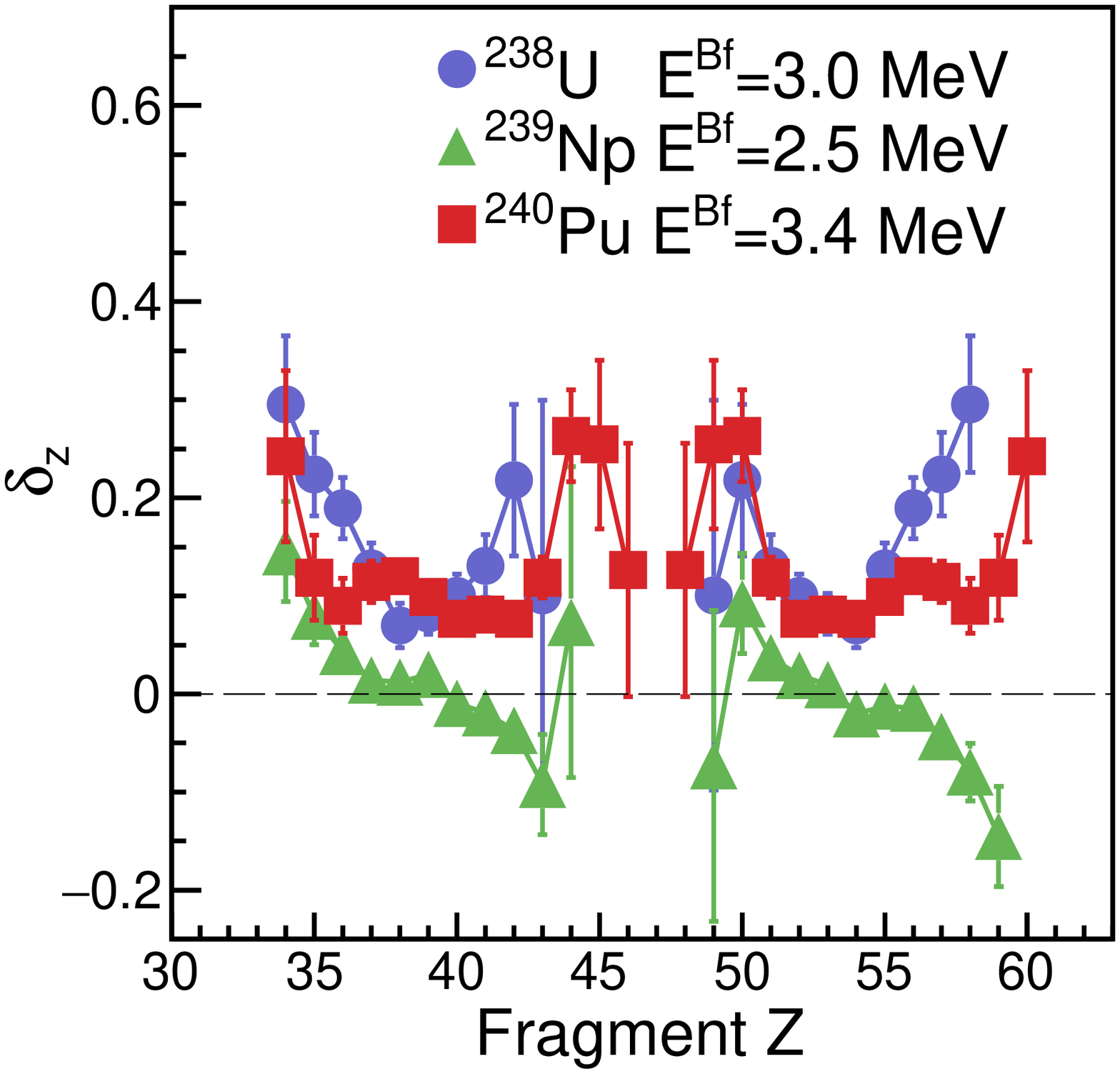}
  \includegraphics[width=0.66\columnwidth]{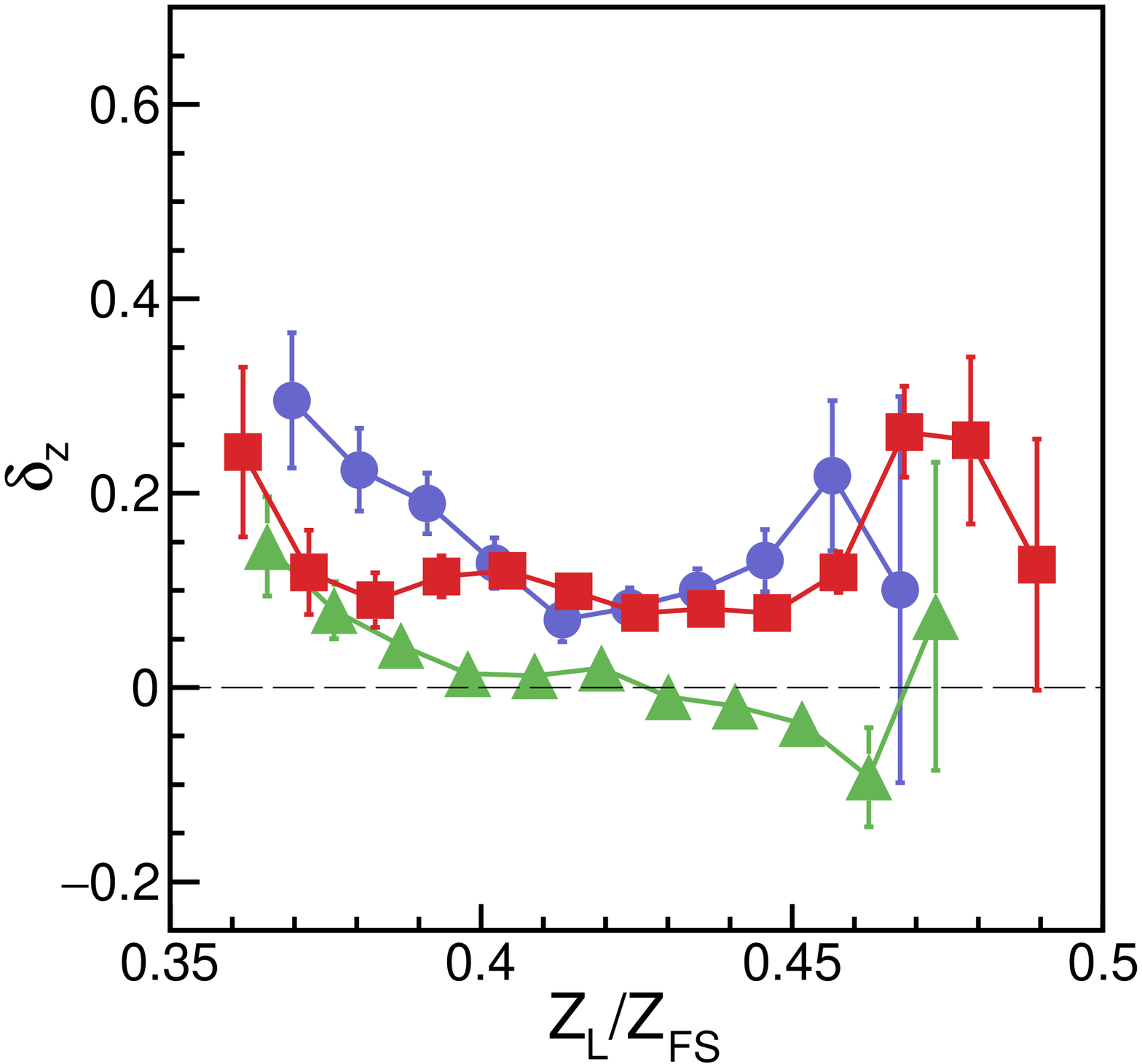}
   \includegraphics[width=0.66\columnwidth]{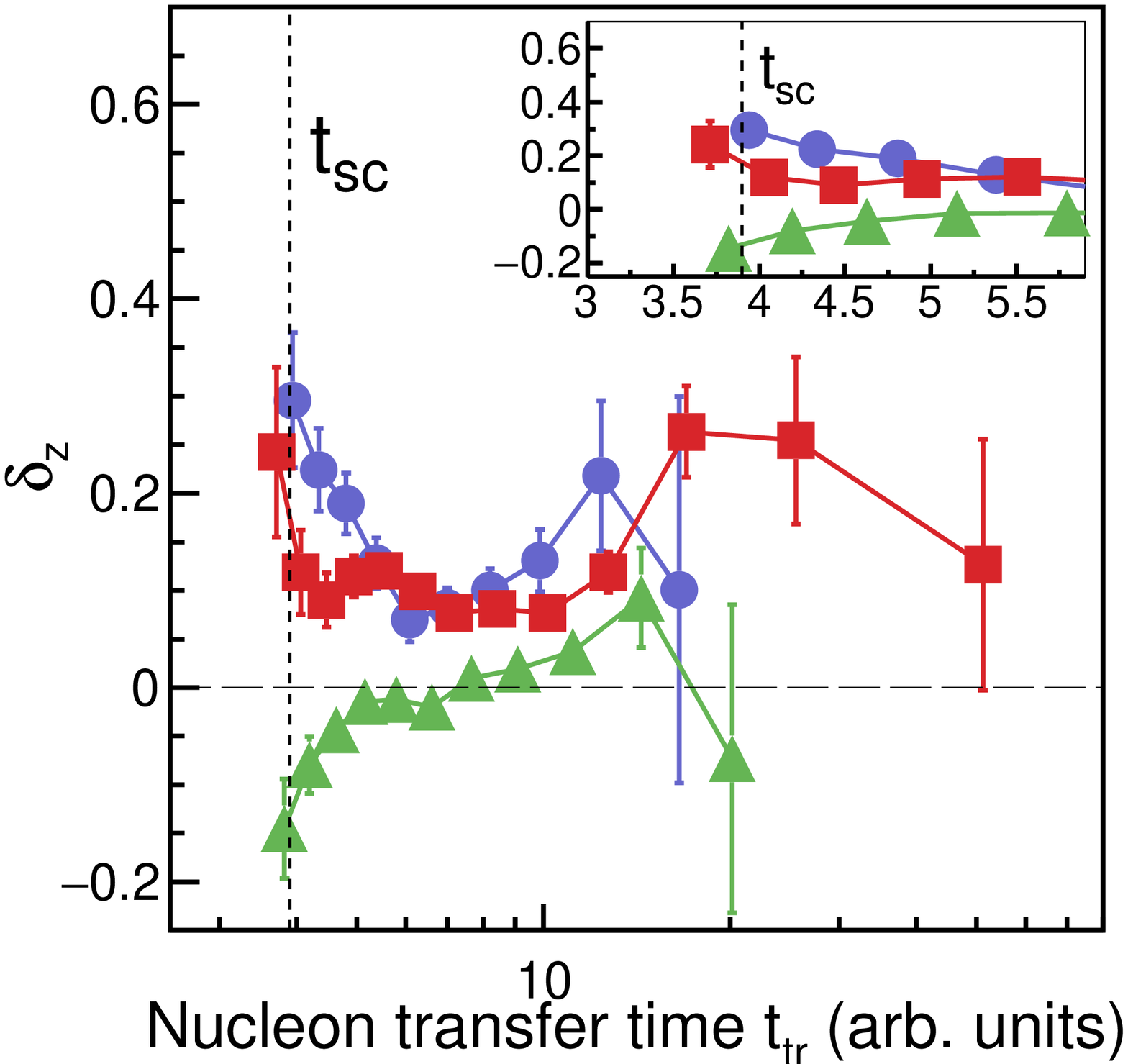}
  \caption{Local even-odd staggering $\delta_{\rm Z}$ as a function of different ordering parameters. In all panels, the points correspond to $^{238}$U at $E^{\rm Bf}$=3.0 MeV (blue dots), $^{239}$Np at $E^{\rm Bf}$=2.5 MeV (green triangles), and $^{240}$Pu at $E^{\rm Bf}$=3.4 MeV (red squares). The panels show $\delta_{\rm Z}$ as a function of the fragment Z (left panel), the asymmetry of the split as the ratio between the light fragment Z and that of the fissioning system (middle panel), and the time to transfer unpaired protons $t_{\rm tr}$ as calculated in Eq.~\ref{eq_tr} (right panel). In this right panel, the vertical dashed line shows the largest asymmetry before threshold splits were identified, and thus where $t_{\rm tr}$ is expected to be similar to $t_{\rm sc}$ (see main text). The inset shows a zoomed area around these points.}
  \label{fig_full_delta}
\end{figure*}

\section{Local even-odd staggering and ordering parameters\label{sec_data}}

By construction, in a fissioning system with Z$_{\rm FS}$ protons, the local even-odd staggering as a function of the fragment Z is symmetric with respect to Z$_{\rm FS}$/2 in the case of even Z$_{\rm FS}$ and antisymmetric in the case of odd Z$_{\rm FS}$. Left panel of Fig.~\ref{fig_full_delta} shows the complete distributions for $^{238}$U, $^{239}$Np, and $^{240}$Pu for similar initial fission energy. The peak around Z=50, which can be clearly seen in all systems, appears at different light-fragment Z$_{\rm L}$ depending on Z$_{\rm FS}$: Z$_{\rm L}$=42, 43 (at negative $\delta_{\rm Z}$), and 44 for $^{238}$U, $^{239}$Np, and $^{240}$Pu, respectively. 

In previous studies of the influence on $\delta_{\rm Z}$ of the relative level density of the fragments, the asymmetry of the split was used as ordering parameter for comparing different systems~\cite{steinNPA,caaJPG11,jurJPG}. Among the different ways that the asymmetry can be expressed, the middle panel of Fig.~\ref{fig_full_delta} shows the measured $\delta_{\rm Z}$ as a function of the Z$_{\rm L}$/Z$_{\rm FS}$ ratio.

\subsection{Scission and proton-transfer times}

As it is discussed in the main text, the average time to transfer all unpaired protons from the light fragment to the heavy one, $t_{\rm tr}$, depends on the energy stored and the difference of temperature $\Delta T$ between them (a discussion can be found in ref.~\cite{schPP}):
\begin{equation}
t_{\rm tr}\propto\frac{E^{\rm int}_{\rm L}}{\Delta T}=\frac{E^{\rm int}_{\rm L}}{T_{\rm L}-T_{\rm H}}.
\label{eq_tr}
\end{equation}
In this frame, the temperature is defined as the inverse slope of the logarithmic of the constant-temperature density formula and it can be expressed with an empirical relation with the fragment mass and shell corrections $S({\rm A,Z})$~\cite{egiPRC09}:
\begin{equation}
T={\rm A}^{-2/3}(0.0570+0.00193~S({\rm A,Z}))^{-1}
\label{eq_temp}
\end{equation}

The right panel in Fig.~\ref{fig_full_delta} shows $\delta_{\rm Z}$ as a function of $t_{\rm tr}$ for the three systems. The dashed line separates the points where the threshold asymmetry is identified. These points correspond to splits where the average transfer time of unpaired protons from the light to the heavy fragment $t_{\rm tr}$ is much shorter than the average scission time $t_{\rm sc}$ (see main text). Thus, one can expect that, right before these points, $t_{\rm tr} \approx t_{\rm sc}$.


\end{document}